\documentclass[10pt,twocolumn,letterpaper]{article}

\usepackage{cvpr}
\usepackage{times}
\usepackage{epsfig}
\usepackage{graphicx}
\usepackage{amsmath}
\usepackage{amssymb}
\usepackage{adjustbox}
\usepackage[english]{babel}
\usepackage{authblk}
\usepackage{tabularx}
\usepackage{xcolor}
\usepackage{colortbl}

\cvprfinalcopy % *** Uncomment this line for the final submission

\begin{document}
\date{}
\title{Leveraging Intel SGX to Create a Nondisclosure Cryptographic library}

\author{Mohammad Hasanzadeh Mofrad and Adam Lee

       Department of Computer Science
       
       University of Pittsburgh
       
       \tt\small \{mohammad.hmofrad and adamlee\}@pitt.edu}

\maketitle

\begin{abstract}
Enforcing integrity and confidentiality of users' application code and data is a challenging mission that any software developer working on an online production grade service is facing. Since cryptology is not a widely understood subject, people on the cutting edge of research and industry are always seeking for new technologies to naturally expand the security of their programs and systems. Intel Software Guard Extension (Intel SGX) is an Intel technology for developers who are looking to protect their software binaries from plausible attacks using hardware instructions. The Intel SGX puts sensitive code and data into CPU-hardened protected regions called enclaves. In this project we leverage the Intel SGX to produce a secure cryptographic library which keeps the generated keys inside an enclave restricting use and dissemination of confidential cryptographic keys. Using enclaves to store the keys we maintain a small Trusted Computing Base (TCB) where we also perform computation on temporary buffers to and from untrusted application code. As a proof of concept, we implemented hashes and symmetric encryption algorithms inside the enclave where we stored hashes, Initialization Vectors (IVs) and random keys and open sourced the code (https://github.com/hmofrad/CryptoEnclave). 
\end{abstract}

\textit{Keywords: Intel Software guard extension, enclave, cryptography, small trusted computing base.}

\section{Introduction} \label{introduction}

Trusted Execution Environment (TEE) provides an isolated and trusted environment embedded in the device during manufacture. This enables the device's processor and memory to protect user installed applications through hardware isolation. A trusted computing base is the minimum amount of code, usually consisting of hardware, firmware, and other software components, that a user must trust. If there is a bug or vulnerability in the trusted computing base, the entire system will contain a security breach.

Mass processor manufacturers like Intel have built embedded hardware technologies to support TEE implementations. The Intel SGX \cite{mckeen2013innovative, hoekstra2013using, anati2013innovativey} is a set of new hardware instructions that enables applications to run code and safeguard important data such as cryptographic keys from within their own protected execution environment. As a result, the CPU becomes the only exposed surface area to remain vulnerable. The Intel SGX is shipped with a Software Development Kit (SDK), which is a collection of development resources in C/C++ to deliver a production quality application. It supports sealing of data and local attestation between enclaves as well as remote attestation by a remote server to verify enclave identity. The SGX cryptographically hashes the code and data in an enclave which is a protected region of memory and only allows the code within the enclave access its data.

The SGX architecture enables developers to create hardware assisted enclaves. These enclaves are opaque containers to the host Operating System (OS). In this project we aim to create a cryptographic library that performs the cryptographic operations inside an enclave, allowing developers to abstract away their application's dependencies on cryptographic functions and fearlessly store their keys inside the enclave. The enclave encrypts and stores the application's sensitive keys such that the application can store its encrypted data outside of the enclave, without fear of another, potentially malicious, application from reading or tampering that data. As with any professionally written cryptographic library, by abstracting away the cryptographic functions, developers can create applications which are less prone to vulnerabilities because of poorly executed cryptography. More importantly, the SGX enclave provides a safe place to store cryptographic keys. In order for an application to encrypt and decrypt its data, it needs to store or have a process to regenerate the same cryptographic keys. If an application's symmetric key or private key is stored in an untrusted location, the encrypted data is vulnerable. An attacker with access to both the data and the cryptographic keys can read and manipulate the data. Similarly, if an application depends on a process to regenerate a key, an attacker can easily discover this process and create the same key. An SGX enclave seals off the keys such that an attacker will not be able to access it. Since the key will never leave the enclave, an application's data will not be prone to data tampering or manipulation type of attack. Even if an attacker places malicious code into the operating system, the keys will remain protected inside of the enclave because the operating system cannot read an enclave's data. 

Intel SGX instructions build and execute the enclave into a special protected memory region with a restricted entry/exit location, which is defined by the developer. Thus, this prevents data leakage. Since keys can never leave the enclave, applications can only encrypt and decrypt its own data with the enclave. It cannot send encrypted data to another application, either on the same machine or a separate one, because that data would be unreadable.

We implemented our project using the \texttt{intel-sgx-sdk} \cite{linux_sgx} from Intel open source. Also, we borrowed code from LibTomCrypt \cite{libtom} to implement cryptographic algorithms. The current implementation of the project supports the following cryptographic functions:
\begin{enumerate}
\item Secure Hash with 256 bits digest (SHA-256).

\item Keyed-hash Message Authentication Code with SHA-256 cryptographic hash function (HMAC-SHA-256).

\item Advanced Encryption Standard (AES) in Electronic Codebook (ECB) mode which supports three different key lengths including 128, 292, and 256 bits (AEC-ECB).

\item AES in Cipher Block Chaining (CBC) mode which supports three different key lengths including 128, 192, and 256 bits (AEC-CBC).

\end{enumerate} 

However, in our future work, we will create a protocol for two enclaves, either on the same machine or separate, to mutually authenticate each other and to create a shared secret key in order to transmit data over an unsafe channel. Unsafe channel is defined to be the mechanism of transmitting data from one enclave to another that is vulnerable to both passive and active attackers. Since malware can reside in an OS, even two enclaves on the same machine must assume to have an unsafe channel between them.  

The remainder of this report is divided into the following sections.
In Section \ref{tools}, we introduce the SDK and libraries we have used to implement our secure cryptographic library. In Section \ref{ncryptolib}, we provide a technical overview of the implemented nondisclosure cryptographic library. In Section \ref{futurework}, we present the ideas and benefits for future work on this project. In Section \ref{conclusion}, we summarize our findings and show our achievements from this research project. Finally, in Section \ref{relatedwork} we discuss some related work. 

\section{Tools} \label{tools}
\subsection{Intel SGX SDK}
The SGX programming model \cite{progref} states that an enclave is an isolated region within the application's address space that only code resided within the enclave can access its code and data. Using SGX, an application is armed with a set of instructions that secures its life cycle as a software. After installing an application, first the application launches an enclave. Next, the enclave contacts the service provider to identify the hardware. Then, upon successful verification the service provider establishes a secure channel to enclave. After that, the enclave starts using the hardware-based encryption to peform cryptographic operations on its data. Finally, in the case of software upgrade, the software can request seal keys to unseal the older versions of the data.

Intel SGX \cite{progref} ships with 17 new instructions for supervisor and user mode that can be categorized into followings:

\begin{enumerate}
\item Enclave build/teardown to allocate/deallocate protected memory for the enclave (ECREATE, EADD, EEXTEND, EINIT, and EREMOVE operations).

\item Enclave entry/exit to enter/exit an enclave (EENTER, ERESUME, EEXIT, and AEX operations).

\item Paging instructions to secure page movement and replacement (EPA, ELDB/U, EWB, EBLOCK, and ETRACK operations).

\item Debug instructions to debug enclaves (EDBGRD, and EDBGWR operations).

\item Enclave security operations to let the enclave prove its identity to an external party (EREPORT, EGETKEY operations).
\end{enumerate}

\subsection{Cryptographic Library}
Because there are a plethora of cryptographic libraries already implemented in C or C++, we did not need to write our own implementation of cryptographic functions. Instead, we first narrowed down our search to reputable, well written, libraries. The first requirement for the library was that it had to be modular. We needed a library that we could break apart and only take a few of the functions that we needed. Our next, and more important, requirement was that the library is secure.  Secure, as defined in this context, means the library cannot be vulnerable to timing attacks, back doors, or weak key generation. A cryptographic algorithm with a large key size is still weak if the implementation is vulnerable to side channel attacks. Our search led us to three candidates that all support symmetric key algorithms, hash functions, pseudo random number generators, and public key algorithms. Three cryptographic libraries that we looked into are:

\begin{enumerate}
\item The first candidate was LibTomCrypt \cite{libtom} library which is open sourced and built to be highly modular, two necessities for our project. While the library does not support SSL or verification of certificates, it contains all the necessary tools to do so. The API was built to be able to support any new cipher, hash, or pseudo random number generator. Thus, as new algorithms are created and old algorithms are broken, very few lines of code will have to be changed to add new, up to date algorithms.
\item The second candidate was OpenSSL \cite{openssl}, written in the C programming language, is very powerful and provides support for Transport Layer Security (TLS) and Secure Socket Layer (SSL) protocols \cite{ssl}. Started in 1998, OpenSSL has been rigorously tested and patched over the years. Since the second phase of our project includes two SGX enclaves sharing keys, the TLS and SSL protocols in OpenSSL would be valuable resources. However, OpenSSL contains much more than this project requires and a more modular library option will likely be chosen. It is worth note that the Github repository of Openssl \cite{opensslgit} has 3,264 stars and 1,866 forks.
\item The third candidate was Crypto++ \cite{cryptopp}, written in the C++ programming language. Crypto++ has documentation for each algorithm in its library and its own Wiki page \cite{cryptoppwiki}. If the project required an entire cryptographic library, Crypto++ would probably be the top choice, since there is an easy installation process. However, the source code for Crypto++ is poorly organized on Github \cite{cryptoppgit}. Since it was written to be shipped all together, there are no sub-directories in the libraries file system structure. This will make it more difficult to take only the algorithms needed for this project.
\end{enumerate}

Because of the complexity of OpenSSL \cite{openssl} and limitations of the \cite{cryptopp} which are stated the above, we chose LibTomCrypt \cite{libtomcrypt} which is a part of an open source project \cite{libtom}. Written in C, LibTomCrypt is well organized and maintained project with 405 stars and 144 forks on its Github repository \cite{libtomcryptgitt}. On its website, LibTomCrypt proclaims to be written to be both modular and portable. While LibTomCrypt is not as widely used as OpenSSL, it has still be implemented in both academic and professional settings. Furthermore, LibTomCrypt is compatible with GCC and Visual C++, which is also compatible with SGX SDK. 

\section{Nondisclosure Cryptographic Library} \label{ncryptolib}
In this project, we select some parts of LibTomCrypt \cite{libtom} cryptographic library and make them compatible with the \texttt{linux-sgx-sdk}  \cite{linux_sgx} from Intel Open Source group (01 dot org). The \texttt{linux-sgx-sdk} is the SGX's Linux SDK written in C and C++, which can be run in both emulation and hardware modes. The project is hosted in Github at https://github.com/01org/linux-sgx. We installed the SDK on an Ubuntu 14.04 LTS 64-bit virtual machine with 2 Virtual CPU, 2GB RAM, and 8GB disk and test our implementation in the simulation mode.

 We created wrappers on the existing cryptographic implementations of LibTomCrypt's cryptographic algorithms. The implemented library is lightweight, modular, and most importantly, secure. Secure, as defined in this context, means the library is not vulnerable to timing attacks, back doors, or weak key generation. Using enclaves as a safe strongbox for keys, our implementation acts as an Application Program Interface (API) for the implemented cryptographic functions.

In the reminder of this chapter we will introduce the threat model and describe the functions that we have implemented and present their implementation details.

\subsection{Threat Model}
In our threat model, we assume a powerful adversary that has complete access to the OS of the machine. This means that the adversary can read all of an application’s data but she cannot read and write into Enclave page Cache (EPC) pages i.e. where the enclave's data is stored. However, we are assuming that the adversary can only perform a remote attack. This means that the adversary does not have physical access to the machine. Moreover, in the case of physical presence of the attacker, we reduced the attack surface to the temporary buffers between an application and its enclave. 

Following Intel SGX's generic threat model, we do not protect against passive address translation attacks where a malicious kernel can obtain the page level trace of an application executing inside the enclave and create a copy of code and data present in the enclave's address space. We also exclude the CPU chip, power analysis, cache timing, and side-channel attacks.

\subsection{Utility functions}
We implemented a generic I/O interface between the application (untrusted code) and enclave (trusted code). In the application part, we have standard C functions like \texttt{open()}, \texttt{read()}, \texttt{close()}, and etc. that are used to open a file and read chunks of it that are passed to the enclave using temporary buffers of size 4KB. We also implemented a command line application that lets a user access different cryptographic algorithms and define arbitrary key lengths for performing the cryptographic operations on input files.

\subsection{Secure Key Generation}
A pseudo random number generator is a key part of any cryptographic library. Keys for HMACs, symmetric keys, Initialization Vectors, and asymmetric keys all rely on random bytes to be generated. A predictable pseudo random number generator with low entropy is a point of attack for an adversary. If an adversary can predict the bytes that are used to generate a key, then the encryption can easily be broken. Even in the asymmetric case, generating keys still requires random numbers. If the attacker can predict the randomness, then she can follow the rest of the key generation process to recompute the private key or reconstruct the plaintext. For our project, we utilized the build in pseudo random number generator that the SGX provides. While LibTomCrypt has a few different algorithms for generating random numbers, using SGX’s implementation was simple. We will leave the incorporation of LibTomCrypt’s pseudo random number generators to give the user options on how their key is generated to our future work. 

As previously stated, one of the goals of this project is to securely generate cryptographic keys inside the enclave. We meet this goal by writing a wrapper on top of the \texttt{sgx\_read\_rand()} \cite{linuxsgxsdk} function which lets the user to generate her arbitrary-sized keys. The \texttt{sgx\_read\_rand()} will only notify the status of operation to the user and the keys will remain inside the enclave. 

\begin{figure}[t]
\begin{center}
\includegraphics[width=\linewidth]{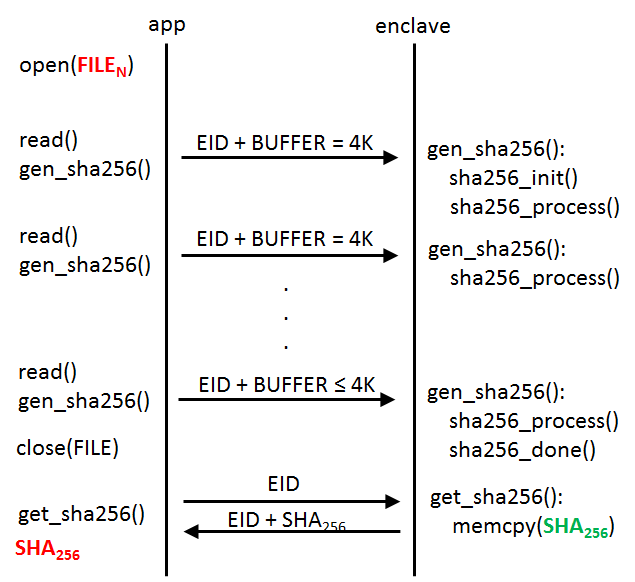}
\end{center}
\caption{A sample SHA-256 scenario: The application is trying to pass temporary buffers to the SHA-256 implementation inside the enclave. The EID stands for Enclave ID.}
\label{fig:sha_256}
\end{figure}

\begin{figure}[t]
\begin{center}
\includegraphics[width=\linewidth]{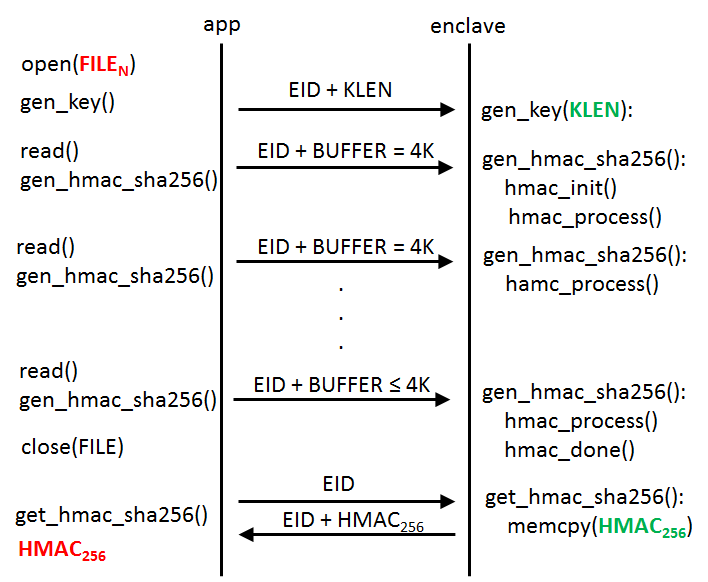}
\end{center}
\caption{A sample SHA-256 scenario: The application is trying to pass temporary buffers to the HMAC-SHA-256 implementation inside the enclave. The EID stands for Enclave ID.}
\label{fig:hmac_sha_256}
\end{figure}

\subsection{Hashes}
We implemented the SHA-256 hash function. We chose SHA-256 because it is the most well-known hash function that does not have any known vulnerabilities. While a hash function does not use any keys and could have been implemented in user space, we needed to implement SHA-256 inside of the enclave in order to implement an HMAC. An HMAC is a keyed hash that is used in many protocols in order to ensure the integrity of data. Assuming the secrecy of the key, HMACs prevent an adversary from modifying data without the user becoming aware of the modification. Thus, it is an important cryptographic primitive that must be implemented inside of the enclave instead of in user space. The followings are some implementation details about the two hash functions:

\begin{enumerate}
\item SHA-256: Figure \ref{fig:sha_256} illustrates our SGX-enabled SHA-256. The user opens a file and then iteratively calls the \texttt{gen\_sha256()} function to build the hash for the input file. All the computations for generating the hash digest out of temporary buffers are done inside the enclave using \texttt{gen\_sha256()} implementation. The user can access enclave functions using the Enclave ID(EID). Finally, the user can retrieve the hash value using \texttt{get\_sha256()}. 

\item HMAC-SHA256: Figure \ref{fig:hmac_sha_256} shows how the SGX-enabled HMAC-SHA-256 works. The HMAC-SHA-256 follows a similar process to SHA-256. First, the user requests the enclave to create a random key by calling \texttt{gen\_key()} which is a wrapper on top of \texttt{sgx\_read\_rand()} function. Then user opens a file and  incrementally calls the \texttt{gen\_hmac\_sha256()} function to build the hash for the input file. The user can access this function by passing the Enclave ID(EID) as a paramter to her function call. The generated key and digest are stored inside the enclave. The user does not have explicit access to the hash result, however, she can access the digest by calling \texttt{get\_hmac\_sha256()} function. 
\end{enumerate}

Regardless of the size of input file, storing the incremental SHA-256 and HMAC-SHA-256 hashes inside the enclave allows us to only store 256 bits digests, which verifies our small TCB requirement.

\begin{figure*}
\begin{center}
\begin{tabular}{cc}
{\includegraphics[height=6cm,keepaspectratio]{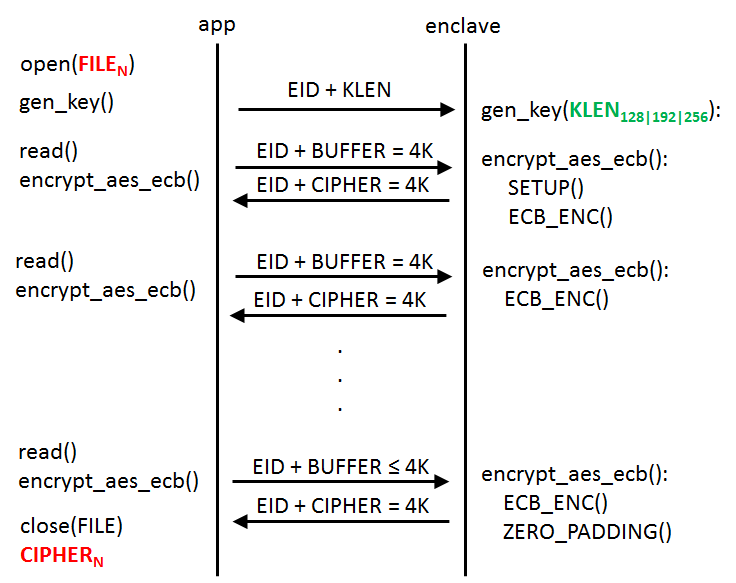}} &
{\includegraphics[height=6cm,keepaspectratio]{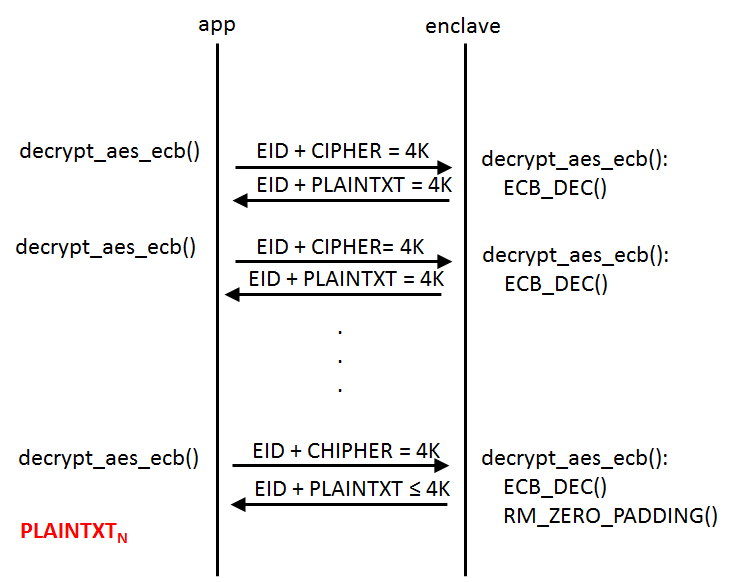}}
\end{tabular}
\end{center}
\caption{A sample AES-ECB scenario: The left side represents the encryption process and the right side represents the decryption process. The EID stands for Enclave ID.}
\label{fig:aes_ecb}
\end{figure*}

\begin{figure*}
\begin{center}
\begin{tabular}{cc}
{\includegraphics[height=6cm,keepaspectratio]{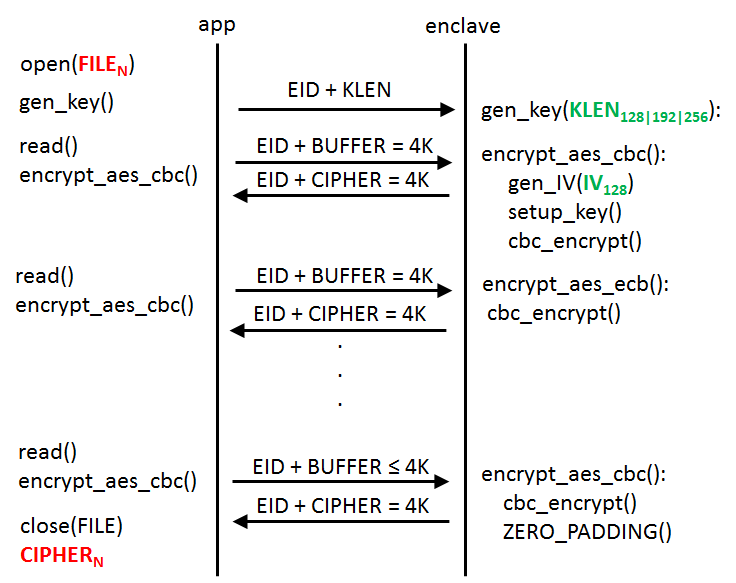}} &
{\includegraphics[height=6cm,keepaspectratio]{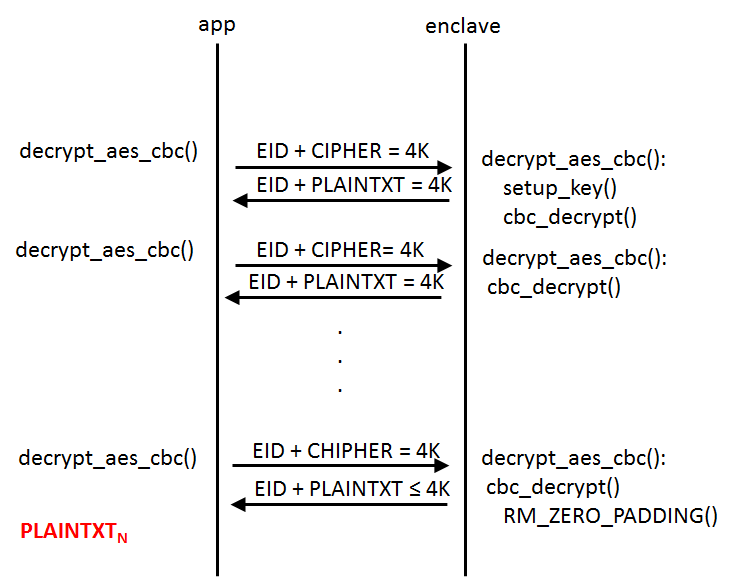}}
\end{tabular}
\end{center}
\caption{A sample AES-CBC scenario: The left side represents the encryption process and the right side represents the decryption process. The EID stands for Enclave ID.}
\label{fig:aes_cbc}
\end{figure*}

\subsection{AES}
We implemented the symmetric key algorithm AES with 128, 192, and 256 bit key sizes. Similar to SHA-256, AES is also the most well-known symmetric key algorithm and is not vulnerable to any known attacks to this date. We first implemented the block chaining mode with Electronic Code Book mode (ECB). While not as secure, this was simpler to implement than the other block chaining modes. Furthermore, we implemented Cipher Block Chaining (CBC) after ECB. While ECB should not be used in many scenarios, we implemented it as an option to the user and because it is a building block for the rest of the block chaining modes in LibTomCrypt library. While LibTomCrypt uses a software based approach for AES, runtime performance gains can be achieved by using the hardware implementation that comes standard to Intel processors. Also, one can implement AES using a library which supports for AES hardware implementation. Shifting this implementation over to hardware approach will be left to future work.The followings are the implementation details for two implemented AES modes:

\begin{enumerate}
\item AES-ECB: Figure \ref{fig:aes_ecb} shows how an application communicates with AES-ECB implementation inside an enclave. In the encryption phase, first the application opens the file and requests a random key of desired size (128, 192, or 256 bits). The enclave will generate the key but the key will remain in the enclave address space. Then the application starts passing temporary buffers of plaintext to the \texttt{encrypt\_aes\_ecb()} function and the function will return the generated ciphertext using the keys stored in the enclave. In the decryption phase, the application contacts the enclave which has the symmetric key and starts passing stored ciphertext to the \texttt{decrypt\_aes\_ecb()} function. Having the key, the enclave will then produce the plaintext and pass it to the application side. Throughout a series of invocations, the application assembles the original plaintext from received buffers and delivers it to the user.

\item AES-CBC: Figure \ref{fig:aes_cbc} shows the overall process of AES algorithm in CBC mode. This process is similar to AES-ECB, except this process requires a 128 bit Initialization Vector (IV). This IV is also generated inside the enclave and stord there. The AES-CBC has two phases: encryption (\texttt{encrypt\_aes\_cbc()}) where temporary buffers of plaintext are passed to the enclave and the generated ciphertext are stored in the application memory, and decryption (\texttt{encrypt\_aes\_cbc()}) where temporary buffers of ciphertext are passed to the enclave and the original plaintext  is generated incrementally inside the application memory using the key and IV stored in the enclave.
\end{enumerate}

Following our principle design decision, we keep the size of our TCB as low as possible while implementing AES. The plaintext and ciphertext are completely stored in the application address space and the enclave just work on the temporary chunks of data. Here, we only store vital information that needs to be shielded from attacks including keys and IVs.

\subsection{Design Decisions}
The implemented I/O interface in the application space works with buffers of size 4KB. We pick 4KB buffers for the following reasons: 

\begin{itemize}
\item The OS supports for 4KB page sizes, thus having larger buffers requires more pages.
\item Since the AES block size is 16 bytes, we need buffers of multiple of 2 to make them compatible with AES blocks. 
\item The enclave heap size is a multiple of 4. Having buffers of multiple of 4 provides us performance advantages in virtual memory. 
\end{itemize}

In our implementation, we tested our prototype using files at most 200MB. We hardcoded the value of buffer size, still we believe our implementation is able to handle buffers of size  1MB. Larger buffers avoid having numerous I/Os and can scale our implementation, but at the same time, we need larger enclaves with larger heap sizes to store these temporary big buffers. In our current implementation the enclave heap size is 100KB, but for handling 1MB buffers we need to initialize enclaves with of size at least 2MB which will increase the size of our TCB 20 times. Still, this would not result to a practical barrier, because to the best of our knowledge SGX-enabled CPUs can initialize enclaves of size 90MB. 

\subsection{Evaluation}
While our current implementation only operates on a per-session basis and keys are not stored in the enclave, our library still operates securely under our threat model. However, even if the keys were stored, our system would still be secure. As we have previously stated, the cryptographic keys are both generated and used for computation inside of our enclaves. Since only the data, not the keys, leave the enclave, an attacker is left with two options. The first option is to break the encryption. Currently, reversing a SHA-256 HMAC or discovering a key to an AES cipher is computationally infeasible, assuming large enough keys are used. Thus, until a vulnerability is found in one of these algorithms, we can assume that a user’s data can be stored safely outside of the enclave. The second option is to break the SGX enclave and read the keys from its memory. In this regard, we are relying on the security of the SGX. Any compromise found in an SGX processor would be a compromise to our system. The only attacks that the SGX is known to be weak against are attacks where the adversary has access to the hardware of the machine e.g. CPU chip, power analysis, cache timing, and side-channel attacks. Since we have assumed a remote attacker in our threat model, an attack to the hardware of the machine is out of the scope of our project. Thus, under the assumed threat model we can conclude that our system can securely store cryptographic keys such that an attacker cannot read an application’s encrypted data. 

The proposed library ships with a rich command line and it is available at (\emph{https://goo.gl/x7cduK}).

\texttt{./app -a}

\texttt{\textless sha256|hmac\textunderscore sha256|aes\textunderscore ecb|aes\textunderscore cbc\textgreater}

\texttt{[-userkey|-randomkey \textless key|keylen\textgreater]}

\texttt{-intext|-infile \textless input\textgreater }

\section{Future Work} \label{futurework}
With limited time in the semester, there is much that must be left to future work. The current uses of our cryptographic library are very limited since we did not implement saving the keys in the enclave. Since it has already been proven that secrets can be saved and recovered inside of an enclave, this was not our focus for the project. However, to make this into a useful application, that would be the first step. The followings are some interesting avenues for future work of this project:

\begin{itemize}
\item Conducting a set of throughput experiments for having a performance comparison between user space implementation of the selected cryptographic algorithms and the SGX-enabled implementation in terms of the expected runtime. Furthermore, we can design an experiment to demonstrate how buffer size impacts the overall performance of the cryptographic library. Laatly, we can also compare the performance and features of our library with a TPM chip and try to observe the performance difference between these two prototypes.

\item To increase the usability of this project, the library should be expanded. Our work should be viewed as a proof of concept. Adding different types of hash functions and symmetric key algorithms would give users more options on how they wanted to encrypt and verify their data. However, it would be more helpful to add other types of cryptographic algorithms such as public key algorithms like RSA, Elliptic Curve Cryptography, and Diffie Hellman key exchanges as well as various pseudo random number generating algorithms. The pseudo random number generator algorithms would allow a user to customize the way her keys are generated, instead of relying on SGX’s implementation of a pseudo random number generator.  

\item The LibTomCrypt \cite{libtomcrypt} does not support hardware instructions for Advanced Encryption Standard New Instructions (AES-NI) and the AES is implemented using software operations. We can look into other open source cryptographic libraries like \cite{mbedtls} and implement the AES in hardware mode in order to improve the performance of AES encryption and decryption.

\item In our implementation, we expose an API to the user which acts as a wrapper on top of some cryptographic functions from the LibTomCrypt \cite{libtomcrypt}. Another approach to provide access to our implemented library inside the enclave would be to modify the LibTomCrypt functions in a way make them compatible to work with user's temporary buffers. Thus, we can grant finer level of access to the users and remove the wrapper layer from our implementation which trivially results a performance gain.

\item Creating a key exchange between two enclaves would make this a much more useful application. Currently, the library can only be used to encrypt and decrypt an application’s own data. However, by creating a protocol to allow two separate enclaves to safely exchange keys over an unsafe network, the library can become a person’s primary means of storing and utilizing cryptographic keys. The protocol to share secrets between two enclaves could be incorporated into existing protocols that hinge on the use of cryptographic keys. In this way, the keys used in these protocols can be secured inside an enclave and protected from software based threats. Furthermore, using SGX’s properties of remote attestation, the enclaves can verify that both users are storing their keys inside of an enclave. This will ensure that both parties will not leak the shared secret key. 

\end{itemize}

\section{Conclusion} \label{conclusion}
As a CPU manufacturer, hardware is not the only innovation incentive that drives Intel. This time, the Cupertino-based company is trying to revolutionize the software industry by keeping the software developer's critical code and data in a hardware protected memory region. In this project, we leveraged the \texttt{linux-sgx-sdk} to produce a secure cryptographic library that acts as a wrapper atop of an existing cryptographic library \texttt{LibTomCrypt} and supports SHA-256, HMAC-SHA-256, AEC-ECB, and AES-CBC cryptographic operations. In this way, we have added a protection layer to the typical cryptographic operations by using hardware protected enclaves to preserve user privacy and prevent malicious applications from accessing the generated keys. The keys generated using this cryptographic library will never leave the enclave and the user who requests a key through this library will never have direct access to the key. In essence, we implement a secure cryptographic API as a service protected through hardware to be safe from any untrusted software or hardware.

{\small
\bibliographystyle{unsrt}
\bibliographystyle{ieee}
\bibliography{main}
}

\section{Related Work} \label{relatedwork}
There are three articles from three different Intel's research groups  dated back to 2013 that introduce the Intel SGX extension for the first time in a workshop on hardware and architectural support for security and privacy (HASP).

The first paper is written by Frank McKeen.et al. \cite{mckeen2013innovative} introduced the concept of an enclave within an application's virtual address space and showed how these hardware enforceable containers enable application to execute with confidentiality and integrity. They gave a flavor of instruction set and software model of SGX along with its data structures and components. Comparing with a technical report like an Intel manual, they fairly described SGX instructions to load, create, enter, and exit an enclave and enclave paging.

The second paper from the initial series of SGX publications \cite{hoekstra2013using} discusses the usage of SGX in protecting enterprise rights and data as a technological solution that is helping developers to ensure that even naive users can safely manage their personal, financial, and organizational properties without limiting their user experience. Also, they introduced the fundamental SGX terminology. The most significant terms include enclave, measurement, and attestation. An enclave is an isolated region of memory. measurement describes the cryptographic hash of the code and data within an enclave. Lastly, attestation is the mechanism to verify an enclave's identity. Finally, they illustrated how the SGX fits in the software life cycle of an enterprise using One-time Password (OTP), Enterprise Right Management (ERM), and Secure Video Conferencing (SVC).

From their illustrations, we talk a little more about OTP. OTP is an authentication technology often used as a second factor to authenticate users. It is a one-time password solution to authorize online financial transactions. Implementing an OTP within an enclave is somehow similar to our idea of implementing a crypto system inside an enclave. Here the OTP prototype prevents malicious software from gaining access to the OTP pre-shared keys which is exactly our goal in this project.

The third paper from the primary research track of SGX \cite{anati2013innovativey} described the components that allow secure remote provision of enclaves over the network. They proposed a new software lifecycle for shipping software without sensitive data. This process consists of launching an enclave, remote attestation for requesting sensitive data from the service provider, data provisioning from the service provider using a secure channel, data sealing/unsealing to securely encrypt and store data in the enclave, and finally software upgrade for migrating data between different versions of the software. 

The mechanisms that they have proposed to allow secure attestation and sealing provides a mean for the enclave to prove its identity to a remote service provider. In our context, we require to ship keys across the network via secure remote cryptographic operations. So, we can follow the implemented attestation and sealing policies of SGX in order to create a tamper resistant key encryption/decryption over the network.

The second wave of SGX's publications from Intel Corporation belongs to SGX2 which was publicly introduced in 2016. In this research paper, F. McKeen et al. \cite{mckeen2016intel} extends the SGX instruction set to support dynamic memory management for enclaves which removes some limitations of SGX1. The SGX2 addresses three shortcomings of SGX1. First, in SGX1, all enclave memory must be allocated at the enclave's build time, which slows the build time. Also, in SGX1 enclave's memory size cannot expand or shrink while an application is running on an enclave. Second, in SGX1 the access permissions associated with an enclave page are stored in the Enclave Page Cache Map (EPCM) which cannot be changed after a page is added to an enclave. This means that a developer should set highest required permissions during an enclave's build time to allow all possible operations on the enclave. Third, SGX1 does not completely provide library OS support for secure exception handling and lazy loading code inside an enclave. To rectify these shortcomings, in SGX2 the memory management function is split between the OS memory manager (external manager) and the internal enclave resource manager (internal manager).

In summary, the SGX2 instructions enable better protection for code and data inside an enclave. Also, these instructions added support for dynamic memory management (Enclave \texttt{malloc} and \texttt{free}), multithreading, and changing the EPCM's permissions in runtime that empowers an enclave to escalate and deescalate the page permissions for garbage collection. Also, the SGX2's library OSes provide a new type of container where an application is bundled with an OS runtime that executes in the user level. Finally, these library OSes provide secure exception handling for enclaves. The new Intel's SGX2 Instruction Set Architecture (ISA) brings a handful of exciting new hallmarks for system software developers. Unfortunately, after a quick surf inside the Intel's developer zone, we found out that the current hardware does not support SGX2 instructions, however, SGX1 instructions should suffice for porting a cryptographic library into an enclave because it provides the basic required functionality to protect the keys from being leaked.

Ignoring the emergence of the SGX version 2, Jacob I. Torrey et al. \cite{torrey2016enclaves} from Intel Corporation write about the design decisions that a system developer may consider to protect data on a compromised environment. They reintroduce Hardened Anti-Reverse-Engineering System (HARES), which is a thin hypervisor utilizing the Extended Page Table (EPT). EPT can set certain memory pages to execute-only mode and prevent the data pages from being executed by malicious software. The HARES hypervisor uses on-CPU Advanced Encryption Standard - New Instructions (AES-NI) to encrypt/decrypt memory pages. In the presence of a permission violation, the hardware will trigger a VM exit to the HARES to update the EPT permissions with proper permissions. Comparing to the traditional approaches of protecting application via isolation, HARES shares a great deal of similarities with SGX, like setting page permissions and utilizing CPU registers for storing encryption keys during software runtime.

They present three case studies to show how SGX affords system protection against both software and hardware attacks. The first example is the \textit{BlacPOS/Kaptoxa} which is a malware that caused the Point of Sale (POS) breach of Target Corporation. In order to place a financial transaction, the POS devices sends the Personally Identifiable Information (PII) such as a card number or personal identification number to a server for authorization. They claimed that Intel SGX could have secured the PII memory by creating enclaves which denied access of malware to the PII memory region.

The second case study is the \textit{Bolware/Eupuds} which is malware that performed the Man In The Middle (MITM) attack on Boleto payments in Brazil. There is a security plugin that Boleto's customer should download and install on their browser. The Eupuds malware had the ability to detect this plugin and by performing some \texttt{JMP} instructions bypassed this security layer. The authors claimed that, if they could have utilized enclaves to perform different levels of sanity checks for data packets in the browser and prohibit the malware from modifying the data flow. 

In the third case study, they present examples of cross-process injection which obfuscates malicious code within a trusted code. The \textit{Flame} which blends itself with \texttt{iexplorer.exe} closely mimicking the system behavior and \textit{BetaBot} which injects its malicious code into Avast anti virus are two prime examples of this type of attack. The authors state that if the applications targeted by cross-process injection attack were placed in protected enclaves, this kind of attack would have been prevented. SGX has higher level of privileges than windows kernel functions, and SGX will not allow malicious code to spawn a new process inside an enclave. 

Enclaves typically make use of features provided by the memory management unit (MMU) in addition to encryption to make themselves opaque to malicious intrusion. In our proposal, we try to protect the cryptographic operations inside an enclave. This is similar to following the recommendations that the authors gave the corporations such as Microsoft or Target to secure their products from adversaries. 

Trusted Platform Module (TPM) \cite{arthur2015apractical} introduced the concept of software attestation using an auxiliary on-board chip for commodity computers. Despite its wide deployment, this approach is not used in many security systems due to the ever-increasing pace of software updates and the impossible maintenance overhead. Compare to SGX which is embedded in the CPU chip, TPM is an standalone chip which communicates with the CPU via the communication bus. Therefore, TMP is vulnerable to attackers who can tap the communication bus between the CPU and the TPM. Technically, TPM is not vulnerable to software attacks, yet it is vulnerable to an attacker who has physical access to the machine. Finally, it is impossible to remove a software who is measured by TPM from the measurement stored in the TPM Platform Configuration Register (PCR) because TPM tries to measure and verify the software after each reboot.

TPM is an international standard for a secure cryptographic processor whereas SGX is a set of Intel's proprietary CPU instructions. These two technologies have similar capabilities such as remote attestation, data binding and sealing. In this project, we are going to use SGX instructions to implement a cryptographic library, yet one can accomplish this mission using TPM specification. 

Intel's Trusted Execution Environment (TXT) \cite{wojtczuk2009attacking} uses TPM's software attestation but reduces the software running within the secure container to a Virtual Machine Monitor (VMM). TXT secures the software inside the container by ensuring that the container has exclusive control over the entire machine while it is active. Even though TXT provides basic building blocks of an immune system but it is not designed to provide runtime protection against software attacks like buffer overflow attacks. On the other hand, combining TXT with Intel Virtual Machine Extension (VMX) allows system developers to create secure operating systems. 

Compared to TPM and its transcendant Intel TXT, Intel SGX cannot replace the traditional secure boot technologies. SGX primarily tackles the software development life-cycle instead of BIOS and boot security. The Intel TXT triggered the first move for reducing the software inside the container which recently rises as SGX's enclaves for storing sensitive code and data. Although we will not study the possibility of implementing a protected cryptographic library in the context of TXT, but applying this idea to TXT seems feasible and promising. 

\textit{Intel SGX Explained} was written by V. Costan and S. Devadas \cite{costanintel}, two research scientists from MIT, and it presents an in depth study of SGX. Instead of directly jumping to inspecting the SGX architecture, the authors start the paper with information about computer architecture and security which is a prerequisite for analyzing the SGX. They comprehensively investigate SGX hardware design, implementation details, and programing model based on Intel SGX patents, tutorials, and developer's manuals. Finally, they use this information to analyze SGX security properties and conclude that SGX is not resistant to software side-channel attacks. 

While the hardware behind SGX lacks publicly available documentation, they point out a few avenues for SGX analysis. Based on their observations, SGX is vulnerable to debugging ports and DRAM address line bus tapping attacks because the code and data inside the enclave are stored in plaintext in the on-chip cache. As a result, the enclave contents travel on the uncore's ring bus without any cryptographic protection.

Modern Intel processors come with hyper-threading feature that shares the execution unit and caches on a core by two enabled Logical Processors (LPs). Since the SGX does not prevent hyper-threading, a malicious system software can execute a snooping thread on a core shared with a victim enclave. Then, this snooping thread can learn the enclave's executed instructions and memory access patterns. This problem can be solved simply by disabling the hyper-threading feature or fundamentally
by fixing hyper-threading vulnerability. 

Unfortunately, SGX does not protect against passive address translation attacks. Since the memory accesses issued by a SGX enclave goes through the Intel architecture's address translation process, a malicious kernel can obtain the page level trace of an application executing inside an enclave and create a copy of code and data present in the enclave's address space.

The SGX threat model excludes CPU chip, power analysis, cache timing, and side-channel attacks because of their high cost, complexity and deep needed expertise. However, these kinds of attacks can be mounted by unprivileged software and SGX can be circumvented dealing with them.

On the other hand, SGX offers some security mechanisms. The SGX uses the Enclave Page Cache Map (EPCM) to store each Enclave Page Cache (EPC) page's position in its virtual memory address space. In order to let EPC be over-subscribed, SGX allows system software to evict EPC pages into untrusted DRAM. The contents of the evicted page and associated EPCM metadata are encrypted to protect confidentiality, integrity, and freshness. This approach protects enclaves against active attacks using page swapping.

Also, the SGX design prevents malicious software to read and write the EPC pages i.e where the enclave data is stored because the SGX is implemented in processor's microcode which has higher privilege than the malicious software and also SGX's microcode always regulates switching between enclave and non-enclave codes.

Partner with Intel Corporation, Microsoft Research utilizes SGX for its Haven prototype \cite{baumann2015shielding}. Haven's goal is to shield applications from an untrusted cloud. Andrew Baumann et al. leverage the hardware protection of Intel SGX to defend the unmodified legacy applications such as an SQL server and Apache. These legacy applications are running on a malicious OS, but are protected against privileged code and physical attacks. Haven brings confidentiality and integrity to the code and data running on a compromised OS. Haven runs the entire legacy application in an enclave, which is a bad practice due to creating a specific amount of underutilized memory at enclaves build time and limiting the entire application (trusted or untrusted parts) passing through the SGX instructions. 

The Haven paper \cite{baumann2015shielding} points out three primary deficiencies of SGX. First, they discovered that some of the SGX internal states are exposed to the OS exception handler. Second, SGX does not support dynamic memory allocation where enclave pages cannot be added after creation. Third, Enclave page permissions cannot be changed after creation. As we reported in \cite{mckeen2016intel}, these three shortcomings were recently addressed in SGX version \cite{mckeen2016intel}.

In its threat model, Haven attempts to protect against Iago attacks - a class of attacks which arises under the presence of a malicious OS - by limiting the scope of core OS primitives within an enclave and narrowing the exposed SGX interface to attacks. Haven addresses these requirements by emulating unsupported SGX instructions and adding a set of primitive operations to two Windows 8 APIs, the Library OS and Shield module. The Library OS  implements the Windows 8 API using a small set of primitives such as threads, virtual memory, and file I/O. The Shield module is one part of the TCB which provides the ABI required by Library OS to access a set of core OS operations. The TCB set of Haven is larger than what we are going to trust in our project, since we only trust the enclave itself.

Verifiable Confidential Cloud Computing (VC3) \cite{schuster2015vc3} from Microsoft research is the first system that relies on Intel SGX to secure Hadoop MapReduce Cloud workloads and ensures correctness and completeness. They claimed that the VC3 treat model accounts for adversaries who control the entire cloud infrastructure but at the same page, they move the Denial of Service (DoS), hardware side-channels and fault injection attacks out of the picture.

An interesting design principle of VC3 \cite{schuster2015vc3} is to work on an unmodified version of Hadoop. Also, they reduce the Trusted Computing Base (TCB) from Hadoop and operating system to binding the Hadoop jobs to a small amount of code that implement the VC3 cryptographic protocol. VC3 uploads this code to a secure SGX enclave in each Hadoop's worker node, make it accessible to operating system and runs its key exchange protocol inside it to encrypt/decrypt map and reduce tasks.

The VC3 is implemented using a SGX emulator that contains handler for all primary SGX  instructions. In our case, since we do not have access to an Intel CPU shipped with SGX, we may follow their approach to implement our project. Also, VC3 uses enclaves memory to provide a reliable key exchange mechanism for launching map and reduce jobs. Similar to them, one of our assumptions in our project is to keep the generated keys inside a SGX enclave.

Before official release of SGX emulator from Intel Open Source (01 dot org), the researchers from the Georgia Institute of Technology were busy writing the OpenSGX, an SGX emulator. Launching the official SGX open source kernel module by Intel killed the potential opportunities for OpenSGX, unless these researchers proposed a new practical usage for their SGX implementation. In the OpenSGX paper, Seongmin Kim et al. \cite{kim2015first} investigate the design choices of applications and protocols operating on commodity Trusted Execution Environment (TEE). They explore the possibility of using Intel SGX for a wide range of network applications such as software-deﬁned inter-domain routing, peer-to-peer anonymity networks, and middle-boxes. Also, they measure the potential overhead of the SGX-enabled design by implementing these network applications in the context of OpenSGX. 

In the OpenSGX paper, Seongmin Kim et al. investigate the design choices of applications and protocols operating on commodity Trusted Execution Environment (TEE). They explore the possibility of using Intel SGX for a wide range of network applications such as software-deﬁned inter-domain routing, peer-to-peer anonymity networks, and middle-boxes. Also, they measure the potential overhead of the SGX-enabled design by implementing these network applications in the context of OpenSGX. 

Prerit Jain et al. \cite{jain2016opensgx} from Georgia Institute of Technology argues the fact that TEE seriously lags behind its hardware counterparts and a certain group of researchers has only access to the latest hardware breakthroughs such as Intel SGX. Thus, they reintroduce the OpenSGX, which emulates Intel SGX instructions as an open platform for SGX research.They use QEMU's userspace binary translation to implement SGX instruction. The OpenSGX is not an exclusive instruction emulator, yet can be seen as a tiny OS which is able to load enclaves into programs. Moreover, OpenSGX has a user space API, debugging tool, and performance monitoring tool.

The last OpenSGX's paper \cite{shih2016s} from Georgia Institute of Technology tries to secure the Network Function Virtualization (NFV) applications by providing a secure framework. They use OpenSGX to protect specific NFV states which need protection and move these states inside enclaves. In their threat model, they assume NFV applications are deployed in an untrusted data center. Finally, as a proof of concept they use \textit{Snort} a lightweight Network Intrusion Detection System (NIDS). They demonstrate their work by securing the tag state of Snort which helps the packet processing with both OpenSGX emulator and SGX-equipped machines. Leveraging the remote attestation feature of SGX, the S-NFV further supports secure bootstrap for measuring whether an application is booted correctly and established a secure communication channel i.e. it binds a key pair to the remote attestation protocol.

After launching the official release of Liunx-SGX emulator from Intel Open Source, and shipping the Intel's Skylake processor in 2015 and its successor Kaby Lake processor in 2016 which are equipped with SGX extension, it seems Georgia Tech researches are not of the inclination of using OpenSGX anymore and as of right now they are mostly demonstrating their scenarios and ideas using both OpenSGX and SGX-enabled hardware.

\texttt{IPMe} proposed by Anitha Gollamudi et al. \cite{gollamudi2016automatic} is a novel calculus that shows the necessity of SGX-like enclave mechanism. In this security-typed calculus, they show how SGX enforces confidentiality policies against arbitrary attacks on non-enclave code that can corrupt enclave code. \texttt{IPMe} captures an imperative higher-order calculus that captures the key features of enclaves and supports the specification of information security policies. Since we are focusing on a practical problem and this paper is focused on mathematics, we only needed to understand the broader ideas in this paper.

In \cite{jana2013scanner}, the authors propose DARKLY, a privacy protection system for adding a protection layer between an untrusted software (e.g. a perceptual sensor) running on top of an trusted hardware (e.g. OpenCV as a trusted vision API) to preserve user privacy and prevent malicious applications from accessing the hardware. Their proposed privacy protection solution is based on the statement that most legitimate applications do not need unrestricted access to raw physical resources. They evaluated their proposed layer of security using 20 existing OpenCV applications and they observed no significant degradation in functionality and accuracy of these applications.

One of the interesting byproducts of DARKLY is that almost all of the tested OpenCV applications can connect to DARKLY without any modifications. In words, DARKLY's Application Programming Interface (API) lets unmodified OpenCV applications to run throughout this trusted API in a secure manner. Thus, people can easily secure they applications using DARKLY API.

DARKLY tries to provide privacy protection for perceptual applications and this scheme can be observed in our work as well. We are trying to secure cryptographic operations by placing them into an enclave which is functionally similar to the way DARKLY lets OpenCV applications securely communicate with perceptual hardware like camera, headphone, or etc.

In an attempt to patch the weak isolation in software containers shipping within a multi-tenant environment, Sergei Arnautov et al. introduce SCONE (Secure Container Environment) \cite{arnautov2016scone} which implements a cryptographic handle to encrypt/decrypt data for eliminating the SGX system calls overhead. It also supports user-level threading and asynchronous system calls. The design decisions of SCONE leads to a small TCB which follows the idea of keeping the container size small and low performance overhead.This avoids the transition overhead of enclave's execution state and the high cost of non EPC page faults.

Their threat model considers a powerful adversary which has \textit{superuser} privileges and hardware access. The attacker has access to the system software and OS kernel and can modify network packets. They ignore DoD and side-channel, timing and page fault attacks because exploiting them is expensive for both attacker and system maintainer. 

In order to keep the TCB as small as possible, they observed that most of network services communicates are via network sockets or \texttt{stdin/stdout} streams, so they implement a library to transparently shield the outgoing system calls by encrypting/decrypting their communications per file descriptor. Also, they use Transport Layer Security (TLS) to protect the network communication of enclaves. 

They reduce the SGX overhead by implementing the user-level threading inside an enclave that maps OS threads to logical application threads of an enclave and schedules OS threads between enclave threads while having blocking system calls, thus avoiding the need for enclave threads to exit the enclave. 

An inexperienced user can benefit from SCONE as long as she trusts the container creator. Also, for running a SCONE container, the user need a SGX-enabled CPU, SGX kernel module, and an optional kernel module for asynchronous system call support. It is a good idea to understand the design decisions of SCONE and apply them to our problem, since we are also trying to keep our TCP as small as possible.

Raoul Strackx et al. \cite{strackx2016ariadne} propose a set of additive security primitives for protected-module architectures like Intel SGX to avoid exploiting vulnerabilities while these systems crash, reboot, or lose power. Their additional security measures ensure that 1) a protected state can never be rolled back to a previous stale state, 2) accepting an input, the module must eventually finish processing the input or never start processing it, and 3) an unexpected power loss should never cause the system not to start after the reboot. Following these primitives, they propose a mathematical solution to the \textit{state continuity} problem which lies behind these primitives and implement Ariadne, a library providing state continuous storage (\texttt{libariadne}). They tackle an important quirk of SGX technology where SGX enclaves are destroyed when the system is suspended or hibernated. In their threat model, the enclaves are running on top of an untrusted operating system in which its entire software stack is compromised and the attacker can halt the enclave's execution at any moment in time. They claim that adding their implementation to the existing SGX instruction set will fix the state-continuity problem of SGX.

Nico Weichbrodt et al. \cite{weichbrodt2016asyncshock} show that neutral synchronization bugs in enclave code can turn into severe security flaws that help an attacker to hijack enclave's control flow or bypass access control. In their threat model, they assume that an attacker has full access to the OS and can start and stop enclaves. They propose the AsynchShock tool, a semi-automated tool for thread manipulation that exploits the use-after-free and time-of-check-to-time-of-use (TOCTTOU) synchronization bugs inside an enclave. They show how these bugs can be combined with interrupts which are coming from a malicious OS and create a threat. 

Studying the nature of atomicity-violation bugs, they implemented the AsynchShock which is a shared library that utilizes the scheduling pattern of an SGX enclave to trigger an exploit which consists of a series of thread creation, segmentation faults and timer expiration. An example of exploiting the use-after-free bug is that they use \texttt{objdump} to find where the \texttt{free} function is located in the code page, using AsynchShock's registered signal handler, they manipulate the enclave's page access permissions to trigger a segmentation fault for the \texttt{free} function and then initiate a hijack process to exploit the bug.

Florian Tramer et al. \cite{tramer2015sealed} define a formal model, the \textit{transparent enclave execution} that ensures integrity and authenticity but not confidentiality. They formalize the \textit{Sealed Glass Proof (SGP)} that attests the correct execution of snippet while it is running transparently. They present a set of theorem and their associated proofs to ground SGP and as their case study, they use SGP to show the presence of the SQL injection bug in a sample web login page. 

In their threat model, they also tackle the data leakage problem while having side channels. We believe the obvious drawback of their approach is the transparent execution of enclave which makes enclave's data and secret visible to other processes. 

The LibTomCrypt Developer Manual \cite{libtomcrypt} describes the motivation behind the project and everything a developer needs to know in order to use or build upon the existing library. LibTomCrypt is open sourced and built to be highly modular, two necessities for our project. While the library does not support SSL or verification of certificates, it contains all the necessary tools to do so. The API was built to be able to support any new cipher, hash, or pseudo random number generator. Thus, as new algorithms are created and old algorithms are broken, very few lines of code will have to be changed to create new, up to date enclaves.

The manual describes how to use the API for symmetric block ciphers, hash functions, message authentication codes, pseudo random number generators, RSA and Elliptic Curve public key cryptography, and digital signatures. It provides sample code for each category of algorithms and an explanation on how to implement them in a program. Furthermore, the documentation includes advice on key sizes, thread safety, pseudo random number generators, and buffer overflows. This manual simplifies our process of wrapping the library in an enclave. The sample code provided will be a good starting point for our project. 

The Elliptic Curve Cryptology in Practice \cite{bos2014elliptic} was published as a result of the joint work of researchers from Microsoft, the University of Michigan, and the University of Pennsylvania. The paper reviews how elliptic curve cryptology is used in real systems such as Bitcoin, SSL, TLS, and Austrian Citizen ID cards. 

As a result, the researchers found three vulnerabilities to avoid when utilizing elliptic curve cryptology. The first was to use a NIST standardized curve. Many servers were found that used smaller, and thus easier to crack, elliptic curves. The standardized curves are built to avoid known backdoors into breaking the encryption. Second, an elliptic curve can be broken if weak keys are used. Weak keys were defined to be keys that were generated with a poor pseudo random number generator and using the same keys across distinct instances of virtual machines. Third, using a poor pseudo random number generator or reusing the same nonce for digital signatures can leak the private key. 

\cite{nistcrypto2016} was published for the use of federal employees to ensure their cryptographic mechanisms uphold the system requirements for confidentiality, integrity, and authentication. While the audience of the paper was federal employees, in general developers should follow these guidelines to help create secure applications. Mechanisms, not policies, are discussed in this document. For our project, the important take away from this document is the recommended algorithms and key sizes for symmetric cryptographic algorithms, asymmetric cryptographic algorithms, hash functions, and pseudo random number generators. We only want to include in our enclaves the algorithms that have been rigorously tested and do not have any known vulnerabilities.  Since most of the algorithms supported by LibTomCrypt are recommended by this document, the process of implementing only the strongest algorithms from LibTomCrypt into our enclaves will be simple. 

\cite{canetti2001analysis} begins by describing formal definitions for protocols, sessions, and key exchanges. Next, the paper defines two classes of attack models: the Unauthenticated-Links Adversarial and the Authenticated-Links Adversarial. The Unauthenticated version is much more dangerous. That attacker has the power to listen, delay, prevent, change, or create any messages between two parties. In the Authenticated version, the only difference is that the attacker can only transmit existing messages. He or she can no longer create any message. 

Session key security is defined to be that an attacker does not gain any information about a session key by interacting with the key exchange protocol or by attacking other sessions or other users. A protocol is deemed to be secure if an attacker cannot guess more than half of the bits of the session key. The paper then evaluates a few protocols, such as signed and unsigned Diffie-Hellman, based upon the previously defined attack models and the definition of session key security.

\end{document}